\documentclass[aps,prl,reprint,amsmath,amssymb,showpacs,showkeys,superscriptaddress]{revtex4-1}

\usepackage{graphicx}
\usepackage{hyperref}
\usepackage{nomencl}
\usepackage{graphicx}
\usepackage{natbib}
\usepackage{xcolor}

\newcommand{\Fref}[1]{Fig.~\ref{#1}}
\newcommand{\Eref}[1]{Eq.~(\ref{#1})}

\newcommand{\rd}{\rho}
\newcommand{\dmn}{\mathcal{D}}
\newcommand{\nc}{n^\textrm{c}}
\newcommand{\nt}{n^\textrm{t}}
\newcommand{\nNW}{n^\textrm{NW}}
\newcommand{\ncs}{n^\textrm{cs}}
\newcommand{\el}{\ell}
\newcommand{\ec}{c}
\newcommand{\Wa}[3]{W#1/#2--#3}
\renewcommand{\S}{S_2}
\newcommand\vek[1]{\mathbf{#1}} 
\newcommand{\x}{\vek{x}}
\newcommand{\vfrac}{\phi}
\newcommand{\norm}[1]{\|#1\|}
\newcommand{\id}{d}
\newcommand{\np}{n^\mathrm{\id}}
\newcommand{\iW}{t}

\newcommand{\rS}{\widetilde{S}_2}
\newcommand{\ES}{E}
\newcommand\de[1]{\,{\mathrm d}#1}
\newcommand{\Se}{\widehat{S}_2}
\newcommand{\T}[1]{{\boldsymbol #1}}
\renewcommand{\k}{\T{k}}

\newcommand{\rev}[1]{#1}
\newcommand{\revjn}[1]{#1}

\begin{document}

\title{Compressing Random Microstructures via Stochastic Wang Tilings}

\author{Jan Nov\'{a}k}
\email{novakj@cml.fsv.cvut.cz}
\affiliation{%
  Department of Mechanics,
  Faculty of Civil Engineering,
  Czech Technical University in Prague,
  Th\'{a}kurova 7,
  166 29 Praha,
  Czech Republic
}
\author{Anna Ku\v{c}erov\'{a}}
\email{anicka@cml.fsv.cvut.cz}
\affiliation{%
  Department of Mechanics,
  Faculty of Civil Engineering,
  Czech Technical University in Prague,
  Th\'{a}kurova 7,
  166 29 Praha,
  Czech Republic
}
\author{Jan Zeman}
\email{zemanj@cml.fsv.cvut.cz}
\affiliation{%
  Department of Mechanics,
  Faculty of Civil Engineering,
  Czech Technical University in Prague,
  Th\'{a}kurova 7,
  166 29 Praha,
  Czech Republic
}
\affiliation{%
Centre of Excellence IT4Innovations, V\v{S}B-TU Ostrava, 
17. listopadu 15/2172 708 33 Ostrava-Poruba, Czech Republic
}

\date{\today}

\begin{abstract}
This paper presents a stochastic Wang tiling based technique
to compress or reconstruct disordered microstructures on the 
basis of given spatial statistics. Unlike the existing approaches 
based on a single unit cell, it utilizes a finite set of
tiles assembled by a stochastic tiling algorithm, thereby 
allowing to accurately reproduce long-range orientation orders in a 
computationally efficient manner. Although the basic features of the
method are demonstrated for a two-dimensional particulate suspension, the present
framework is fully extensible to generic multi-dimensional media.
\end{abstract}



\maketitle

In 1961, Hao Wang introduced a tiling concept
based on square dominoes with colored edges permitting their mutual assembly in
a geometrically compatible (hard) manner~\cite{wang1961tiling}. Since
then, his tiles have been the subject of studies in discrete
mathematics~\cite{springerlink:10.1007/BF01418780,grünbaum1986tilings,Culik:1996:ASW:245761.245814}
and found an extensive use in computer graphics~\cite{cohen2003wang}, game
industry~\cite{demaine2007jigsaw}, theory of
quasicrystals~\cite{aristoff2011first} or biology~\cite{winfree1998design}. From
the perspective of this paper, the appealing feature of
Wang tilings is that they can compress and reproduce naturally looking planar
patterns or three-dimensional surfaces by employing only a small number of
distinct
tiles~\cite{glassner2004andrew,grünbaum1986tilings,lagae2008comparison}.
Motivated by this observation, we further explore the
potential of Wang tiles to represent long-range spatial correlations in
disordered microstructures; a problem common to materials
science~\cite{Torquato:2002}, geostatistics~\cite{Chiles:1999:GMS} or image
analysis~\cite{Serra:1982:IAMM}.

In this regard, two closely related applications can be distinguished, namely
the \emph{microstructure
reconstruction}~\cite{yeong1998reconstructing,Capek:2009:SRP,davis2011statistically}
based on given spatial statistics and \emph{microstructure
compression}~\cite{Povirk:1995:IMI,zeman2007random,Lee:2009:TDR} aiming at
efficient representation of materials structure in multi-scale
computations~\cite{Fullwood:2010:MSD}. Our focus is on the latter, since these
procedures usually have the microstructure reconstruction techniques at heart,
hereby covering the common features of both.

To the best of our knowledge, compression algorithms reported to date use a
single cell~(PUC) that is periodically extended to tile the plane in a
deterministic manner~\cite{zeman2007random}. Such structures then inevitably
manifest strong long-range correlations with a period of the PUC dimensions. We
shall demonstrate that these artifacts can be effectively controlled when
utilizing small Wang tile sets~\cite{lagae2008comparison,grünbaum1986tilings},
carefully designed to capture morphological trademarks of compressed media,
combined with \rev{fast} stochastic Cohen-Shade-Hiller-Deussen~(CSHD) tiling
algorithm~\cite{cohen2003wang} for \rev{real-time texture generation}. A
potential of this approach will be demonstrated for equilibrium two-dimensional
particulate suspensions consisting of equi-sized disks of radius $\rd$ uniformly
distributed in a homogeneous matrix, cf.~\cite{Rintoul:1997:RSD}.

\begin{figure}[ht]
  \centering
  \begin{tabular}{c@{\hspace{6mm}}c}
  \raisebox{-0.5mm}{\includegraphics{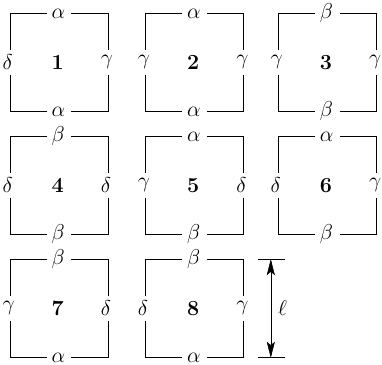}} 
  &\includegraphics{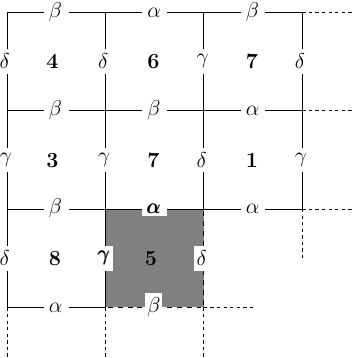} \\
  (a) & (b)
  \end{tabular}
  \caption{\label{fig:0}\rev{(a) A Wang tile set \Wa{$8$}{$2$}{$2$} with edge length $\el$ and codes $\{
    \alpha,\beta,\gamma,\delta \}$.} (b) Example of an aperiodic valid tiling.}
\end{figure}

To this goal, consider a two-dimensional microstructured domain $\dmn$
discretized by a regular square lattice. Each lattice cell contains specific
morphological patterns that are compatible on contiguous boundaries,
\Fref{fig:0}(b). If there are no missing cells in $\dmn$, the discretization is
called a \emph{valid tiling}, and a single cell is referred to as the \emph{Wang
tile}~\cite{wang1961tiling}, \Fref{fig:0}(a). The tiles have different codes on
their edges, lower-case Greek symbols in \Fref{fig:0}(a), and are not allowed to
rotate when tiling a plane. The number of distinct tiles within $\dmn$ is
limited, though arranged in such a fashion that none of them or any of their
sub-sequence periodically repeats. The gathered distinct tiles are referred to
as the \emph{tile set}, \Fref{fig:0}(a). Sets that enable uncountably many,
always aperiodic, tilings are called
\emph{aperiodic}~\cite{grünbaum1986tilings,Culik:1996:ASW:245761.245814}. In
real world applications, the assumption of strict aperiodicity of the tile sets
is relaxed to aperiodicity of tilings, ensured e.g. by the CSHD
algorithm~\cite{cohen2003wang} introduced next.

Intuitively, the ability of a tile set to control long-range order effects
arises from tile and edge code diversities, \Fref{fig:0}(a). Both factors are
related, so that while the number of edge codes $\nc_i$ in $i$--th spatial
direction can be chosen arbitrarily, the number of tiles $\nt$ must
satisfy \rev{
\begin{equation}
  \nt = \nNW \sqrt{\ncs},
  \label{eq:nt}
\end{equation}
}%
where $\ncs = (\nc_1 \nc_2)^2$ is the number of tiles in the
complete set and $\nNW = 2, \ldots, \sqrt{\ncs}$ stands for the optional number
of tiles with identical arrangement of north-western~(NW) edge codes. The
complete set of $\ncs$ tiles is obtained by permuting the chosen codes $\ec_i$.
In valid tilings, the south-eastern edge codes must match those assigned to NW
edges,~\Fref{fig:0}(b). Thus, the tiles in the complete set are sorted according
to existing NW edge code combinations and the desired number of tiles in a user
defined set is formed by selecting $\nNW$ of unique tiles from each group. Such
a set is referred to as \Wa{$\nt$}{$\nc_1$}{$\nc_2$} in what follows. Notice
that the \Wa{$1$}{$1$}{$1$} set corresponds to the PUC setting.

In the stochastic tiling algorithm, the index of a new tile to be placed is
selected randomly with the uniform probability from an appropriate NW group
compatible with the eastern code of the tile previously placed and the southern
code of the tile above the one to be placed (edges $\gamma$ and $\alpha$ in
bold adjacent to shaded cell in~\Fref{fig:0}(b)). Aperiodicity of the resulting
tiling is guaranteed provided that the random generator never returns a periodic
sequence of numbers and that each NW group contains at least two distinct
tiles~\cite{cohen2003wang}.

\begin{figure}[b]
  \centering
  \begin{tabular}{cc}
    \raisebox{3mm}{\includegraphics[height=30mm]{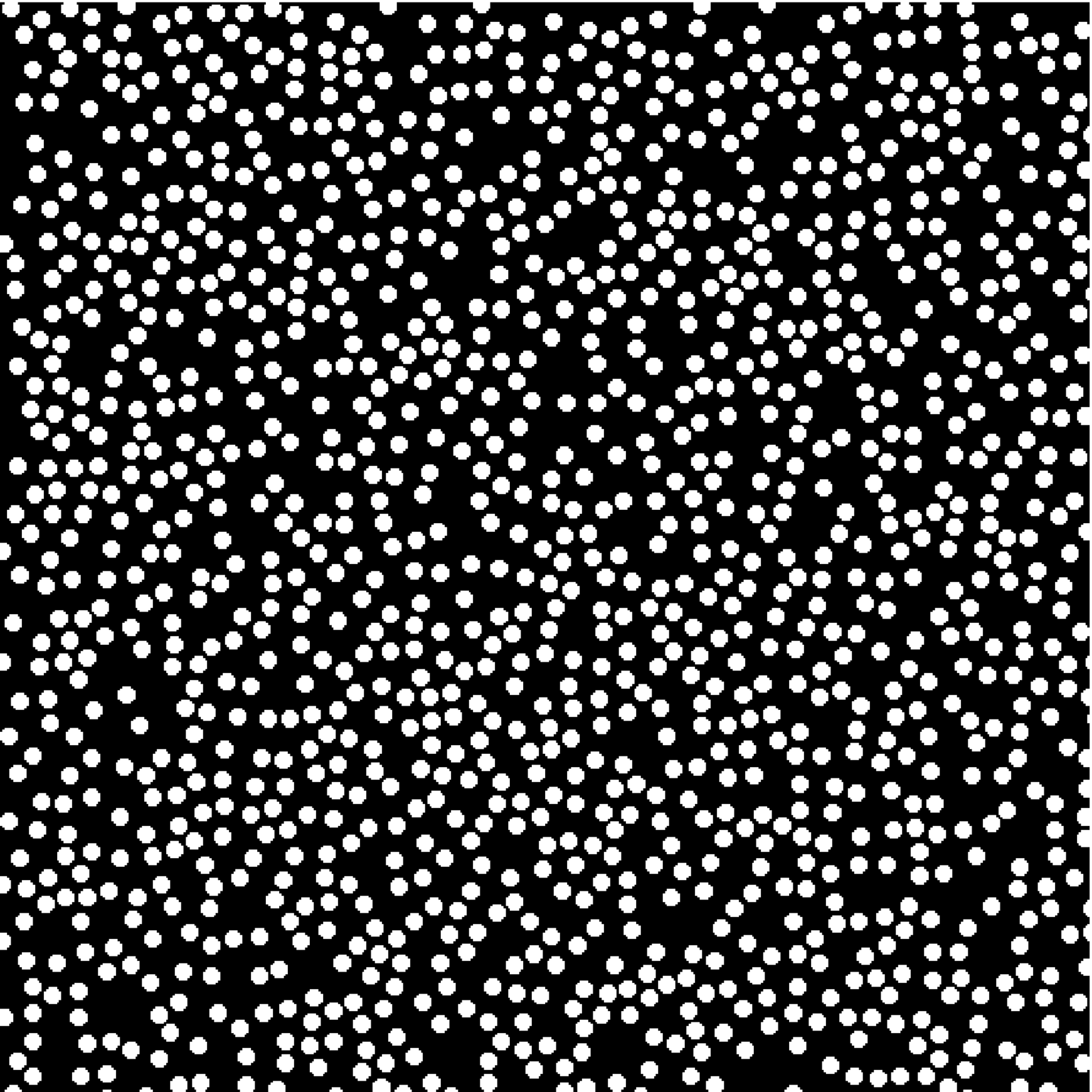}}
    &
    \includegraphics[height=35mm]{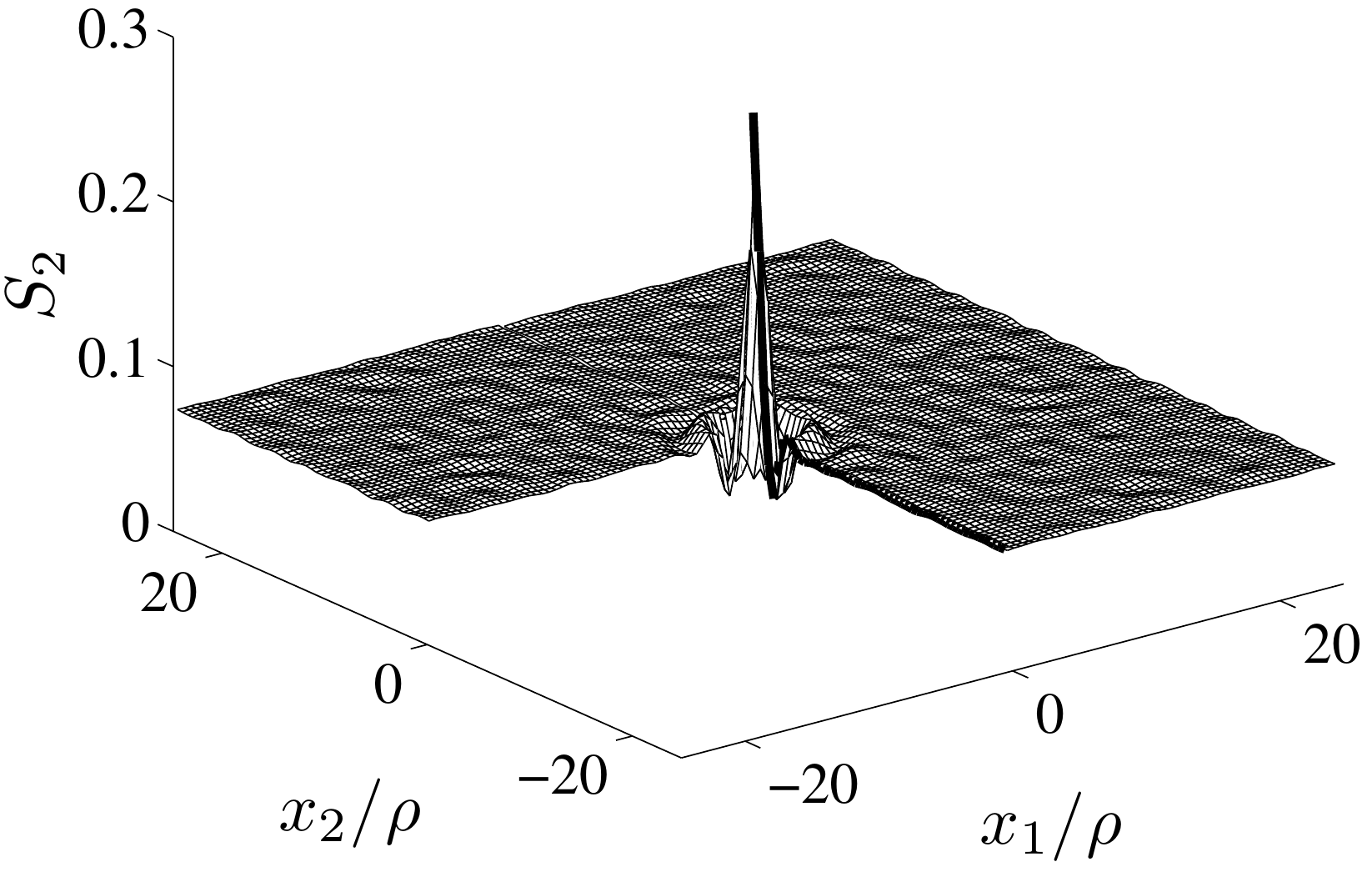}\\
    (a) &(b) 
    \end{tabular}
  \caption{(a) Reference two-phase medium of size $174\rd\times 174\rd$ formed
  by equilibrium distribution of $1,300$ equi-sized disks of volume fraction
  $26.8 \%$ and (b) its two-point probability function $\S$; $\rd$ is the disk
  radius.}
  \label{fig:original_microstructure} 
\end{figure}

Analogously to the existing works on reconstruction and compression of random
media, the tile morphologies are designed by an optimization procedure expressed in
terms of suitable statistical descriptors. As our focus is to control long-range
artifacts, we limit the exposition to the two-point probability function
$\S(\x)$~\cite{Torquato:2002}. For statistically uniform ergodic media, it
provides the probability that two arbitrary points from $\dmn$, separated by
$\x$, are simultaneously found in the particle phase. The function satisfies
$\S(\vek{0})=\vfrac$, where $\vfrac$ is the volume fraction of particles, and
$\S(\x) \approx \vfrac^2$ for $\norm{\x} > \lambda$ indicates the absence of
long-range orders at the characteristic
length~$\lambda$, \Fref{fig:original_microstructure}(b).

\begin{figure}[h]
  \begin{tabular}{cp{4mm}c}
    \includegraphics[height=1.50cm]{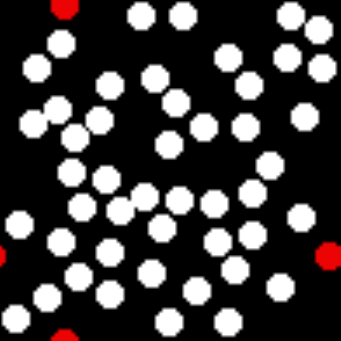}
    &&
    \includegraphics[height=1.10cm]{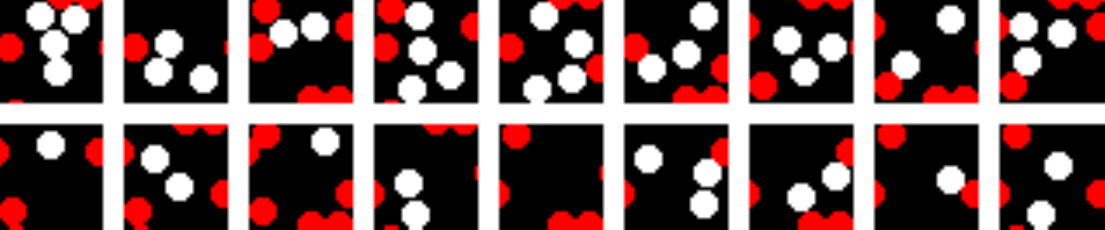}\\
    (a) & &(b)
  \end{tabular}
  \caption{Building blocks of microstructure compression based on (a) PUC
  and (b) tile set \Wa{18}{3}{3} with $49$ disks assigned to tile edges (grey)
  and interiors (white).}
  \label{fig:puc_vs_set}
\end{figure}

In the current setting, the Wang tiling compression consists of a set of $\nt$
tiles of the edge length $\el$, in which we distribute $\np$ disks. The
configuration of particles is determined by the parameter vector $\left[\iW_\id,
\xi_{1,\id}, \xi_{2,\id} \right]_{\id=1}^{\np}$, where $\iW_\id \in \{1, \ldots,
\nt \}$ specifies the parent tile index of the $\id$-th disk and $\xi_{j,\id}
\in [0, \el]$ the local position of the disk at $j$-th direction. To determine
the two-point probability function $\rS$ for a given configuration, we assemble
a tiling that covers the domain of the same size as the representative sample
$\dmn$,~\Fref{fig:original_microstructure}(a). Notice that such tiling
corresponds to a realization of a statistically homogeneous material, since the
tiles are selected from NW edge groups with the uniform probability. The
proximity of the tile-based morphology to the original sample is quantified by
an objective function
\begin{equation}\label{eq:obj_func_tppF}
\ES
=
\frac{1}{|\dmn|}
\int_{\dmn}
\left( 
\S(\x)
-
\rS(\x)
\right)^2\de{\x}
\end{equation}
which can be efficiently evaluated using the Fast Fourier Transform techniques,
e.g.~\cite{Torquato:2002}. The minimization of~\eqref{eq:obj_func_tppF} is
carried out by the Simulated Re-Annealing method \rev{with computational cost
similar to existing PUC design strategies}~\cite{Novak:2011:MEF}. The algorithm
ensures that the tiles in the set satisfy the corner
constraint~\cite{cohen2003wang}, requiring that the tile corners are not
occupied by a disk, and determines the number of disks $\np$ and the cell size
$\el$ such that the local volume fractions associated with edges~(grey disks in
\Fref{fig:puc_vs_set}) and tile interiors~(white disks in \Fref{fig:puc_vs_set})
are as close to the target value $\vfrac$ as possible.

\begin{figure}[b]
  \centering
  \begin{tabular}{cc}
    \raisebox{3mm}{\includegraphics[height=30mm]{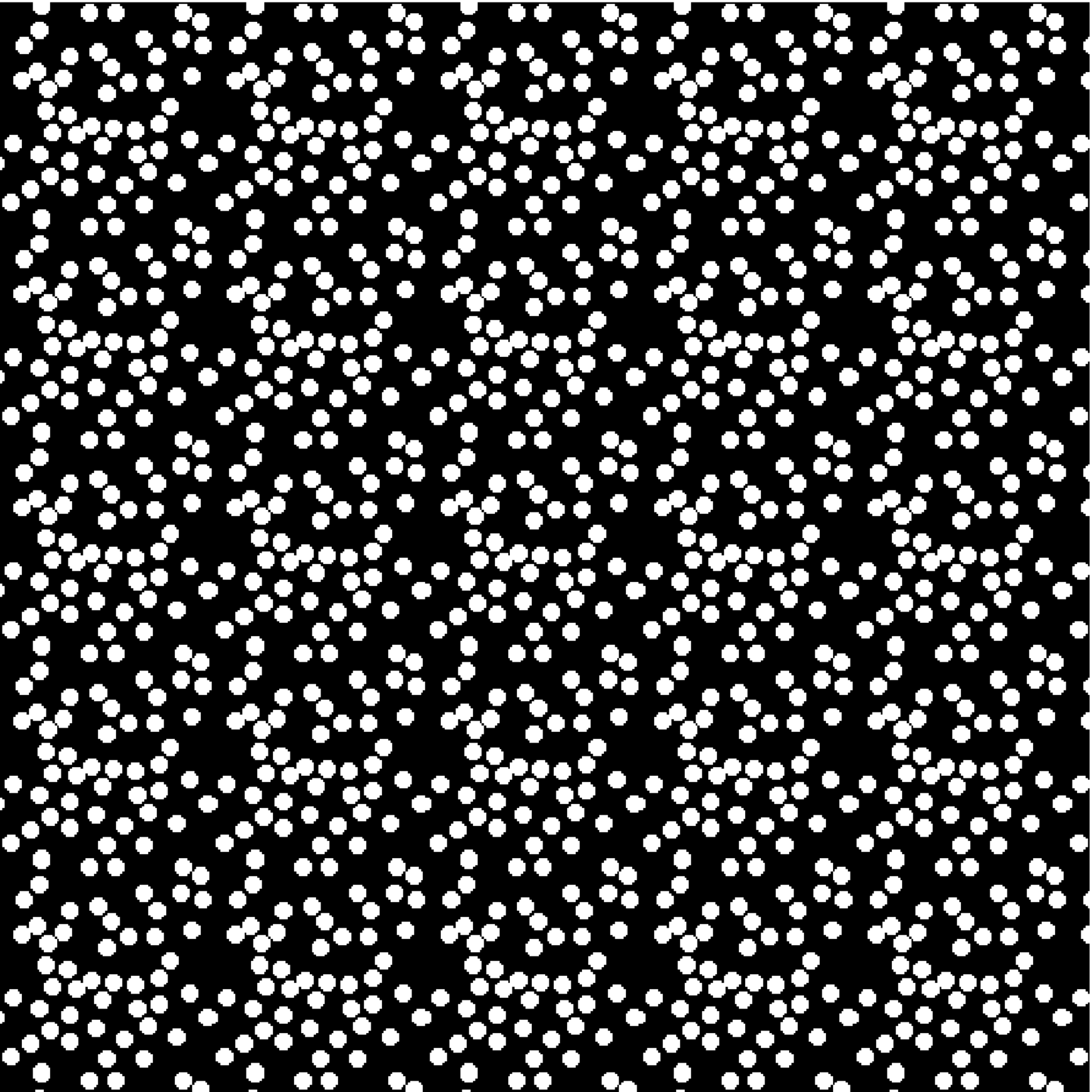}} &
    \includegraphics[height=35mm]{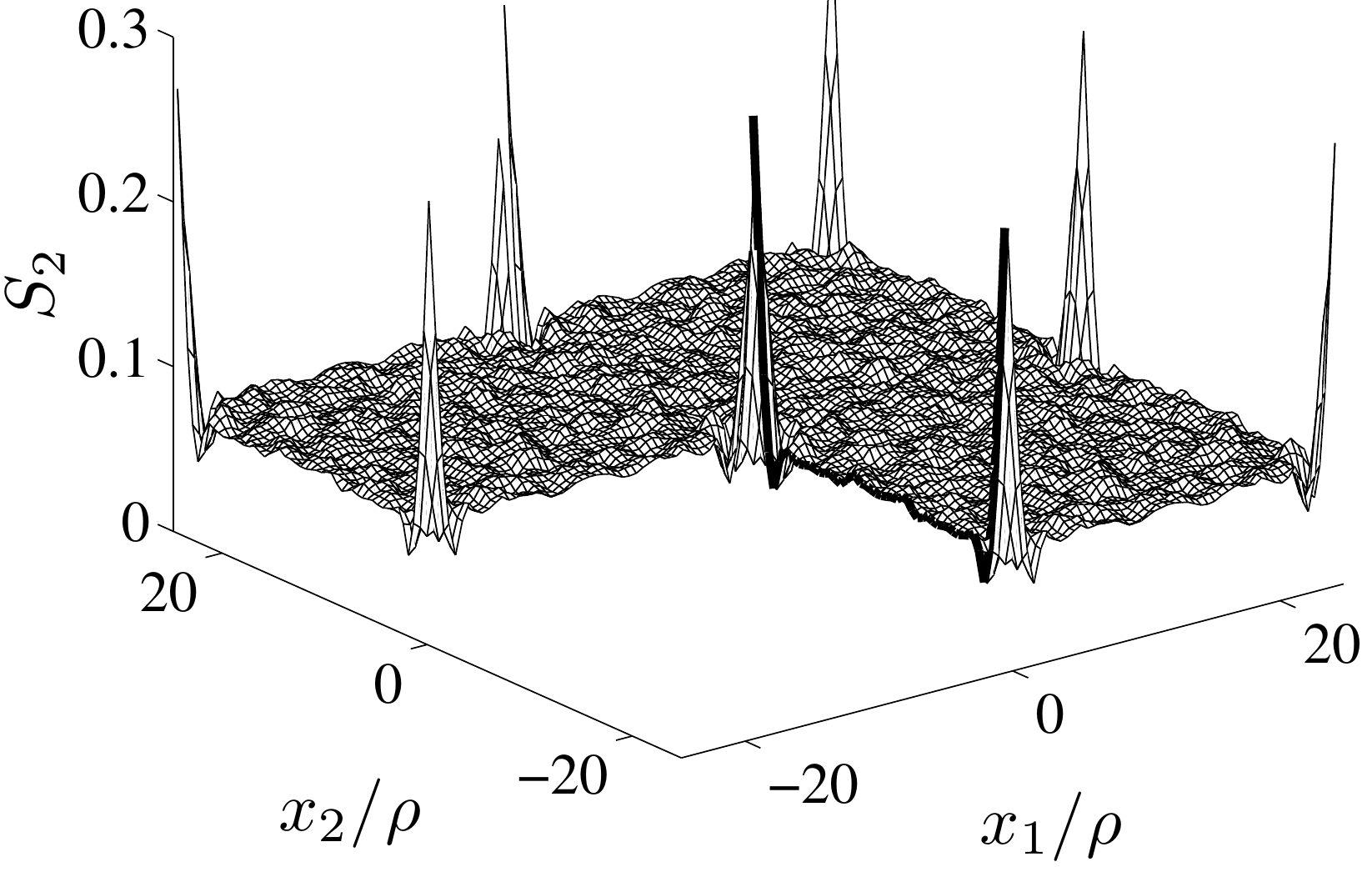} \\
    (a) &(b)\\
    \raisebox{3mm}{\includegraphics[height=30mm]{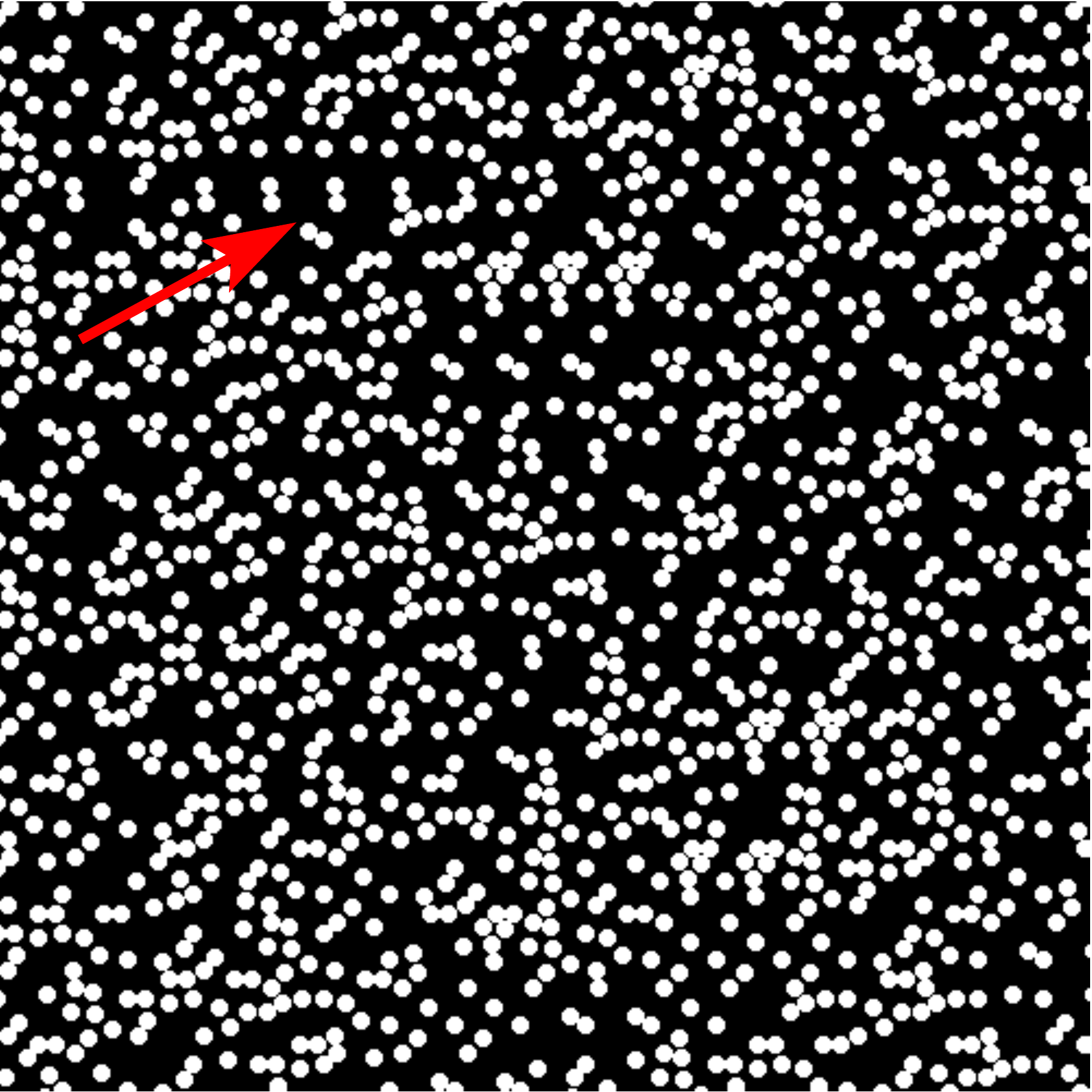}} &
    \includegraphics[height=35mm]{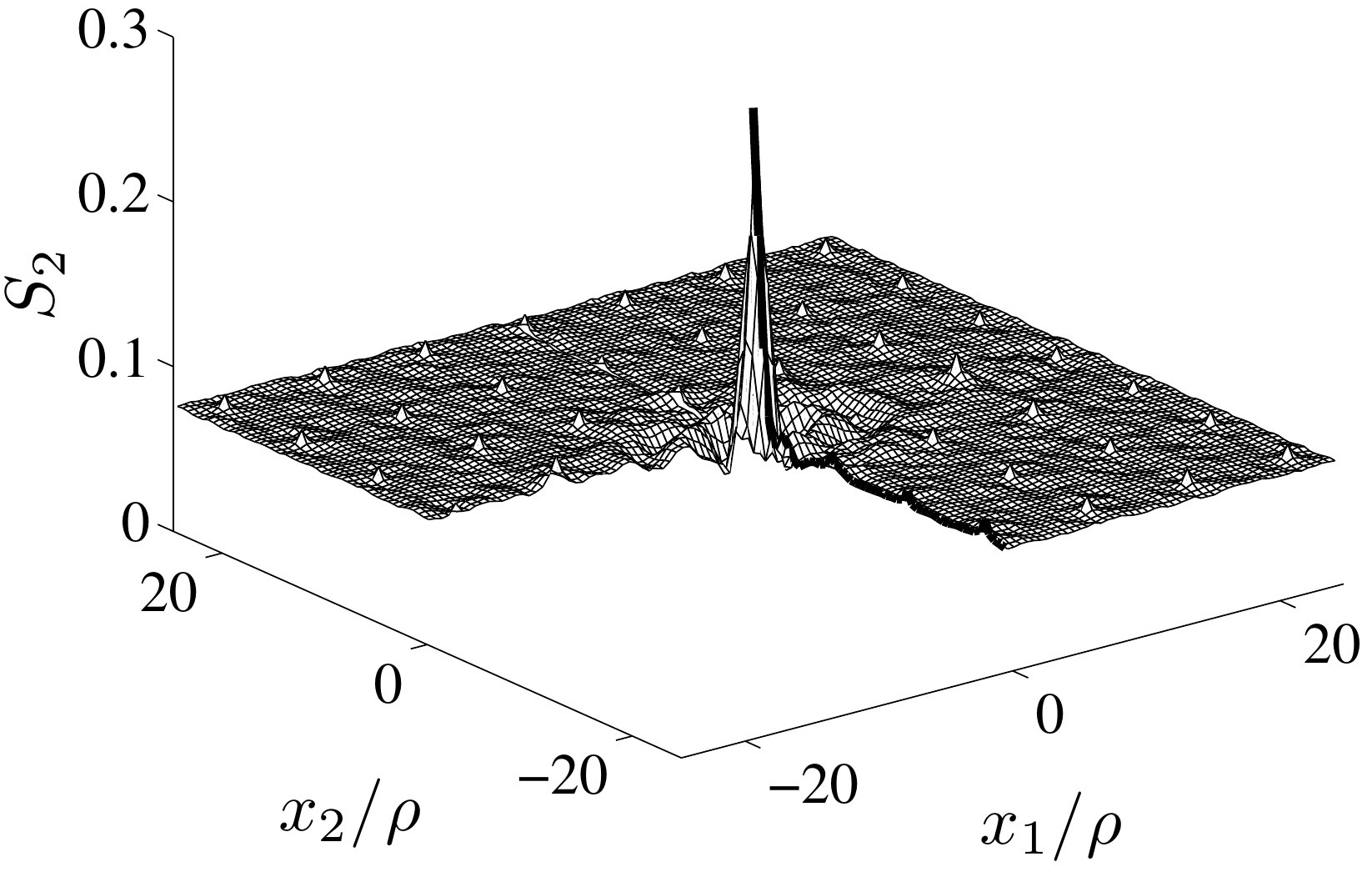} \\ 
    (c) &(d)
    \end{tabular}
  \caption{Optimized microstructures and two point probability functions $\rS$
  for PUC (a,b) and set \Wa{18}{3}{3} (c,d). The arrow in (c) denotes periodic
  region due to local character of tile placement in CSHD tiling algorithm.}
  \label{fig:wangs_vs_PUC}
\end{figure}

\begin{figure*}
\begin{tabular}{ccc}
\includegraphics[height=.18\textheight]{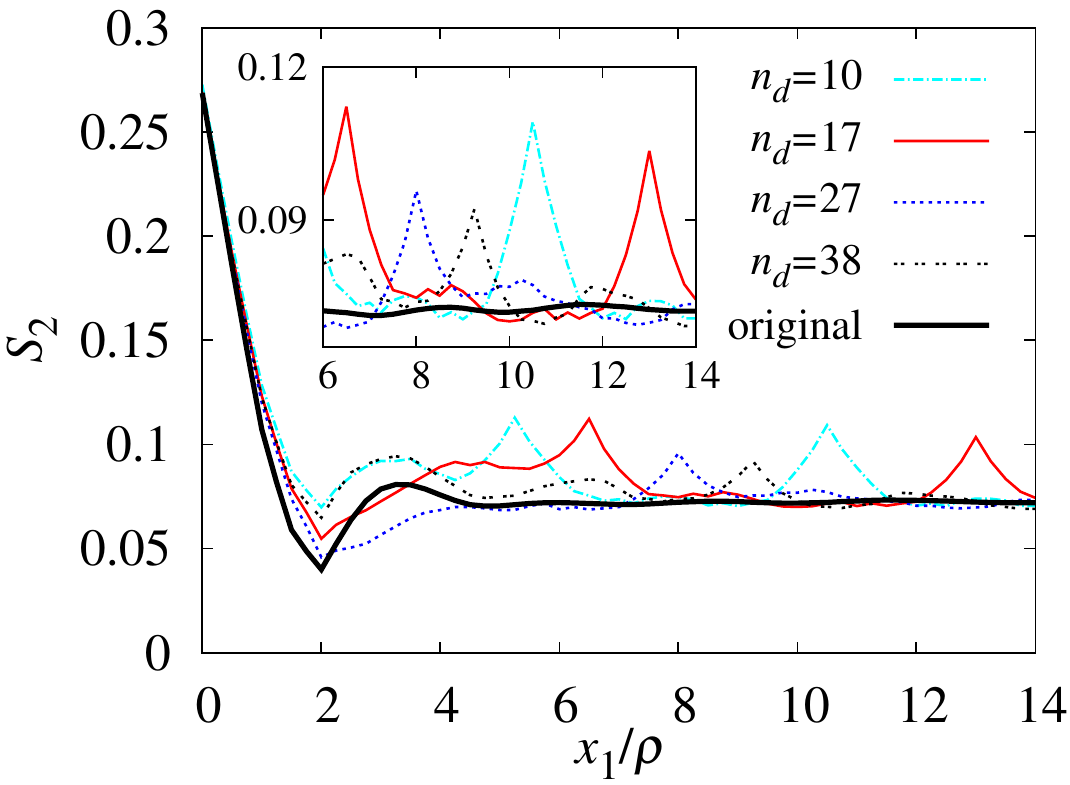} &
\includegraphics[height=.18\textheight]{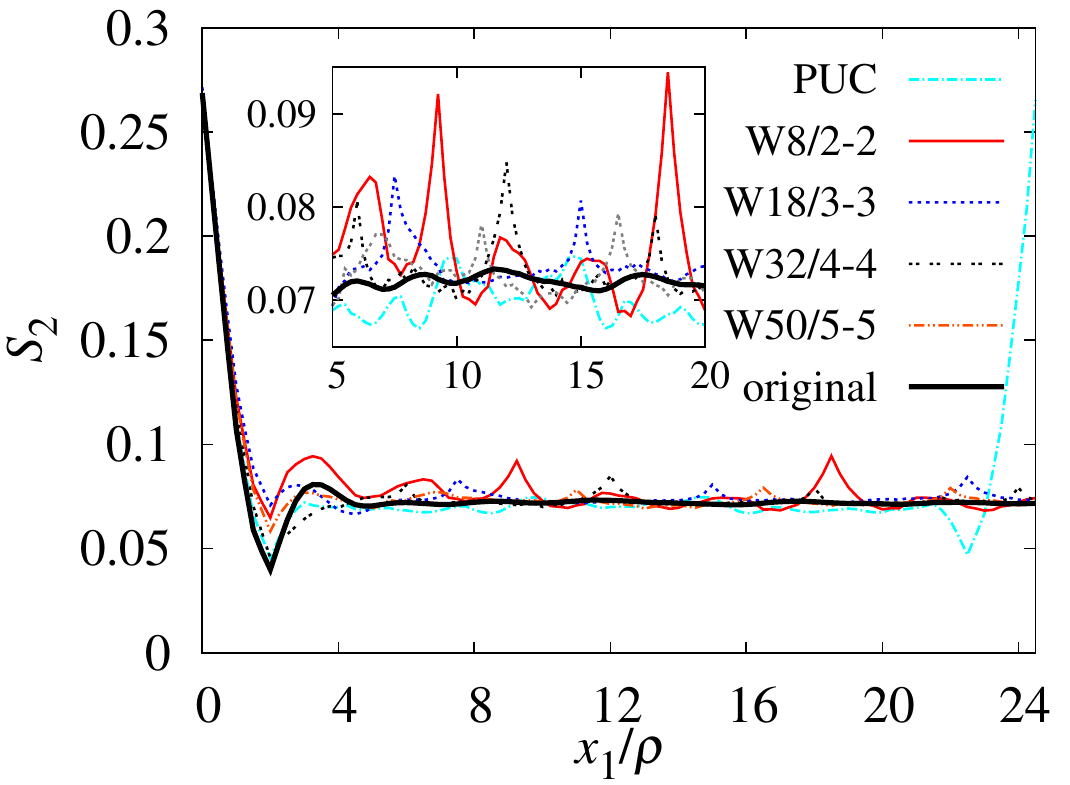} &
\includegraphics[height=.183\textheight]{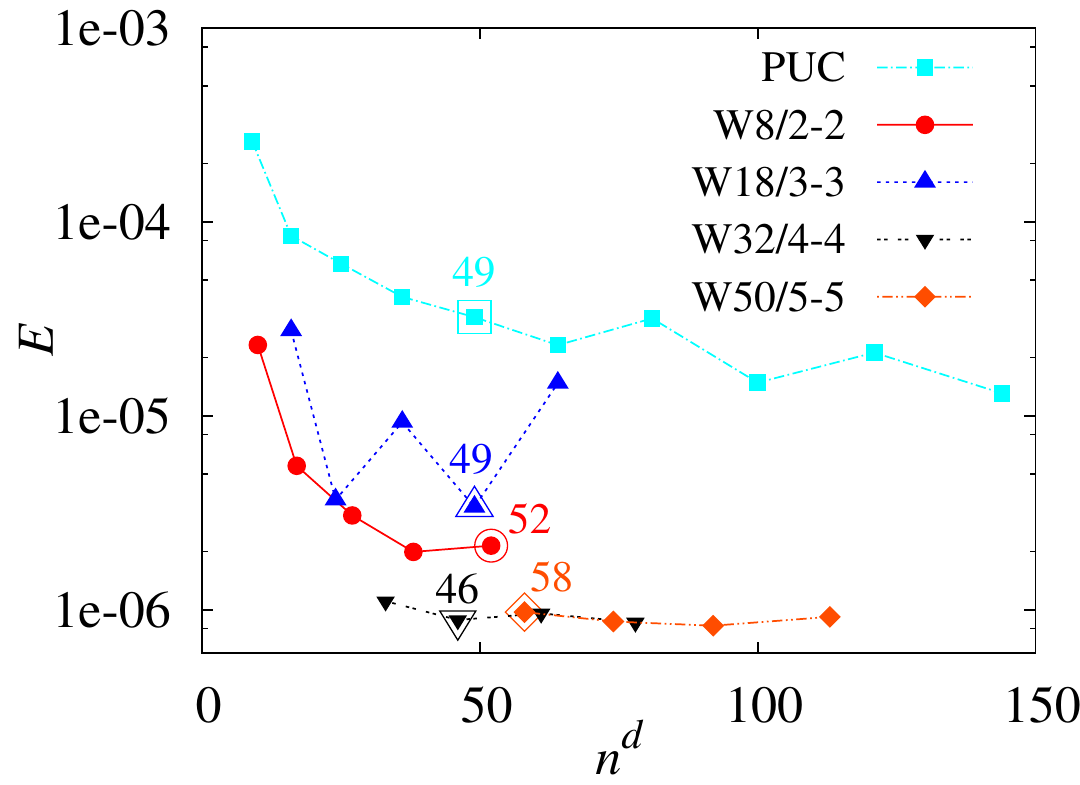} \\
(a) & (b) & (c)
\end{tabular}
\caption{(Color online) Comparison of $\rS$ in $x_1$--$\S$ plane for (a) tile set W8/2-2 with
respect to number of disks $\np$ involved, (b) different sets. (c) Comparison of
different sets in terms of objective function~\eqref{eq:obj_func_tppF} and with
respect to number of disks $\np$ involved. The curves in (b) are plotted for
particular values of $\np$ highlighted in (c).}
\label{fig:acf_comparison}
\end{figure*}

An example of optimal approximations of the target microstructure from
\Fref{fig:original_microstructure} in terms of a PUC and the Wang tile set \Wa{18}{3}{3}
is shown in \Fref{fig:wangs_vs_PUC}. The representations are
based on $\np=49$ particles and tile sizes $\el=24.5\rd$ and $\el=7.5\rd$,
respectively. Evidently, both heterogeneity patterns carry long-range order
effects with the period of $\el$, manifested as the local peaks
\begin{equation}
\Se = \max_{\k \backslash \{ \vek{0} \}} \rS(\k \el)
\end{equation}
in the two-point probability functions, \Fref{fig:wangs_vs_PUC}(b,d). Notice
that $\Se$ is always equal to $\vfrac$ for the PUC approach, whereas the Wang
tiles are capable of adjusting these artifacts by the proper morphology design.
This is also reflected in visual regularity of the generated suspensions,
compare \Fref{fig:original_microstructure}(a) with \Fref{fig:wangs_vs_PUC}(a,c).
Also observe the locally periodic region in \Fref{fig:wangs_vs_PUC}(c), arising
from the local character of CSHD algorithm and from the lowest number of tiles
in groups of admissible NW edge code combinations, $\nNW=2$. \rev{Such
phenomenon is thus less likely when increasing parameter~$\nNW$, however, at the
expense of increasing set sizes, especially for higher edge code diversities,
recall~\Eref{eq:nt}.}

The principal features of Wang tile-based compressions are
further illustrated in \Fref{fig:puc_vs_set}. Instead of relying on a single
cell containing the complete morphological information,
\Fref{fig:puc_vs_set}(a), the tiling-based approach utilizes substantially
simpler building blocks, \Fref{fig:puc_vs_set}(b), assembled to comply with edge
constraints (grey disks in \Fref{fig:puc_vs_set}). \rev{This, however, restricts
the space of admissible disk configurations in Wang tiles compared to the
single PUC design.}

Now, we are in the position to quantify to which extent is the quality of
reconstructed suspensions determined by the tile set diversity and the
morphology design itself. This aspect is examined \rev{first} in
\Fref{fig:acf_comparison} by means of sections of the two-point probabilities
$S_2(x_1,0)$ and the objective function $\ES$, revealing that two effects govern
the amplitude and period of the local extremes $\Se$. First, for a fixed tile
set, increasing the number of disks increases the tile edge dimension $\el$ (and
thus the period) and slightly decreases the amplitude,
\Fref{fig:acf_comparison}(a). On the other hand, increasing the number of tiles
decreases the period as well as the magnitude of local extremes,
\Fref{fig:acf_comparison}(b). Also notice that the quality of the \Wa{18}{3}{3}
set in terms of the objective function~\eqref{eq:obj_func_tppF} is
systematically inferior to \Wa{8}{2}{2}, \Fref{fig:acf_comparison}(c). This is
caused by an inaccurate representation of disk volume fraction for the former
set, which pollutes the shape of $\Se$ statistics,
\Fref{fig:acf_comparison}(b)~\footnote{The best resulting disk volume fraction
for \Wa{18}{3}{3} and $24$ disks was by $1.3\%$ higher than the prescribed
value. The remaining sets, however, never resulted in a worse scatter than
$0.2\%$.}. It further follows from \Fref{fig:acf_comparison}(c) that increasing
the tile set diversity is much more efficient; for sets containing more than
$32$ tiles, the error is almost independent of the number of disks. This
saturation value reflects the inaccurate representation of short-range values of
$S_2$, caused by the particular form of the objective
function~\eqref{eq:obj_func_tppF}. If needed, the local details can be
incorporated \rev{in terms of higher-order statistics} or specifically tailored
descriptors~\cite{yeong1998reconstructing,Capek:2009:SRP,Jiao:2009:SER,Piasecki:2010:EDC}.

It is now clear that the local extremes can be attributed to a limited number of
tiles used in a repetitive, although random fashion. Actually, two components
repeat when tiling the plane: tile edges and interiors. To study the local
artifacts analytically, we consider user defined sets with tiles selected so
that their edges incorporate each code $\ec_i$ at least once. Assuming that
tiles and edges repeat independently, the \revjn{maximum} local extremes can be estimated as
\begin{equation}
\Se^\textrm{p}
\approx
\frac{\vfrac^\mathrm{t}}{\nt}
\left[
\vfrac
+
\left( \nt-1 \right)
\vfrac^2
\right]
+
\displaystyle\max_{i}\left\{
\frac{\vfrac^\mathrm{e}}{\nc_i}
\left[ 
  \vfrac
  +
  (\nc_i - 1)\vfrac^2 
\right]
\right\}
\label{eq:S2_estimate}
\end{equation}
where $\vfrac^\mathrm{t} = (\el - 4\rho)^2 / \el^2$ and $\vfrac^\mathrm{e} = 1 -
\vfrac^\mathrm{t}$ denote the volume fractions of tile interiors~(occupied by
white disks in \Fref{fig:puc_vs_set}) and edges~(occupied by grey disks in
\Fref{fig:puc_vs_set}), respectively~\footnote{%
Observe that the estimate~\eqref{eq:S2_estimate} contains contributions from
tile interiors and edges. In addition, the tile interior part arises from two
complementary events. If the two adjacent tiles are identical, the probability
of simultaneously locating two disks distant by $\el$ amounts to
$\vfrac^\mathrm{t}\vfrac/\nt$. Otherwise, we consider the disks in both tiles
as independent which gives rise to the term $\vfrac^\mathrm{t}(1-1/\nt)\vfrac^2$.
The contribution of repeated tile edges \revjn{related to $i$--th spatial direction} is established analogously, by
estimating the probability of simultaneously matching \revjn{two edges distant by $\el$ as $1/\nc_i$.}}.

\begin{figure}[ht!]
\begin{tabular}{cc}
\includegraphics[height=.17\textheight]{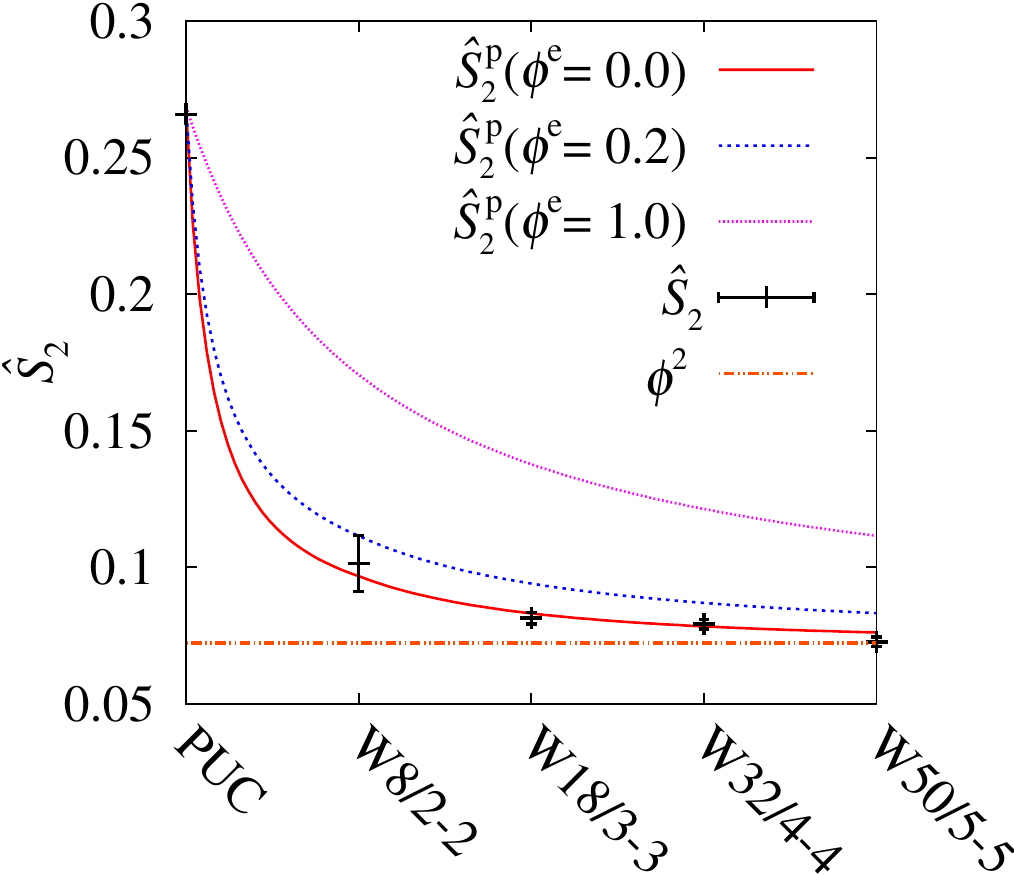} &
\hbox{\hspace{-5mm}\includegraphics[height=.174\textheight]{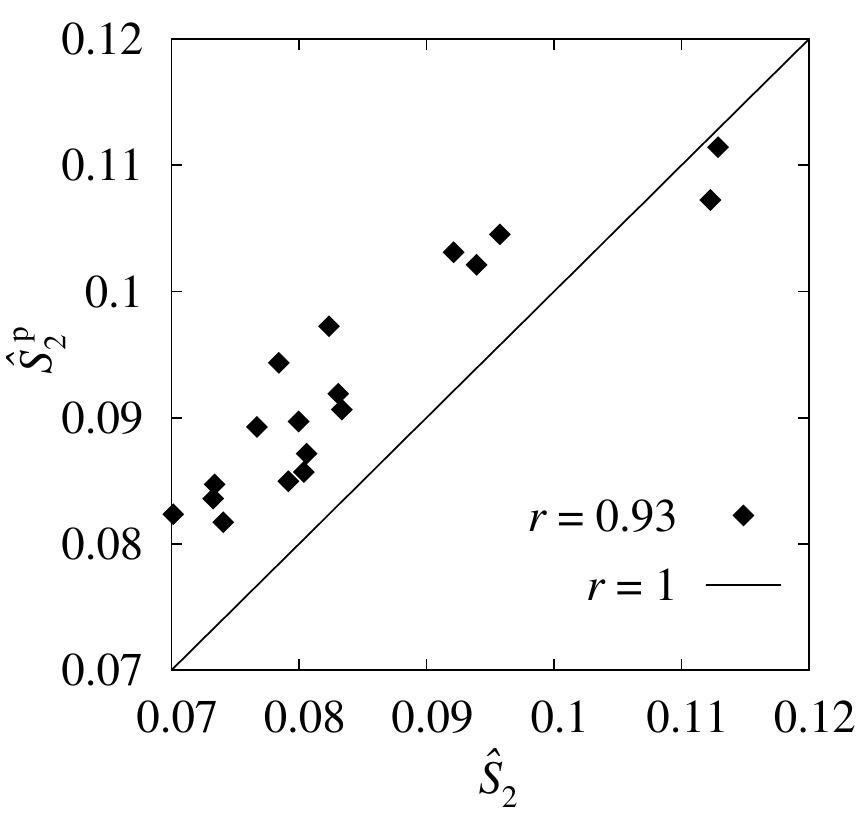}}\\
(a) & (b)
\end{tabular}
\caption{(Color online) (a) Dependence of local extremes $\Se$ on the particular tile
set. (b) Correlation of local peaks obtained from two-point probabilities of optimized
microstructures and their predictions given by \Eref{eq:S2_estimate}; 
$r$ in (b) denotes the Pearson correlation coefficient.}
\label{fig:acf_prediction_comparison}
\end{figure}

\rev{%
In~\Fref{fig:acf_prediction_comparison}, we compare the actual values of $\Se$
with theoretical predictions~\eqref{eq:S2_estimate} for several values of
$\vfrac^\mathrm{e}$. Apart from the limit cases, $\vfrac^\mathrm{e} \in \{0,1\}$,
$\Se^\textrm{p}$ was also explored for $\vfrac^\mathrm{e}=0.2$ (average value
from all considered tile sets). We observe that an almost exact match is
obtained for the lower bound with $\vfrac^\mathrm{e}=0$, red \revjn{solid} curve
in~\Fref{fig:acf_prediction_comparison}(a), demonstrating that the long-range
artifacts are carried mainly by the tile interiors. This is rather surprising,
since all considered tile sets satisfy $\nt \gg \nc_1 = \nc_2$, so that edges
repeat more often than the tile interiors. Moreover, the magnitude of
spatial artifacts converges rapidly to the limit value $\vfrac^2$. Altogether,
this leads us to the conclusion that artifacts due to discrete nature of Wang
tilings can be almost eliminated by a proper morphology optimization. Also note
that the accuracy of the estimate~\eqref{eq:S2_estimate} appears to be
reasonable, both for the average value of $\vfrac^\mathrm{e} = 0.2$, blue \revjn{double dotted} curve
in~\Fref{fig:acf_prediction_comparison}(a), as well as for values corresponding
to individual tile sets, \Fref{fig:acf_prediction_comparison}(b). It may thus
serve as a basis for the a-priori selection of the tile set parameters $\nc_i$ and $\nt$.}

\paragraph{Summary.} A new compression/reconstruction technique based on Wang
tilings has been proposed and applied to two-dimensional microstructures of
disordered particulate media. The technique is extensible to generic
three-dimensional microstructures, adopting the frameworks of Wang
cubes~\cite{Culik:1995:ASW,Lu:2007:VIW} \rev{and image
synthesis~\cite{cohen2003wang}}; it substantially generalizes the periodic unit
cell concept by making use of multiple tiles instead of a single cell, preserves
long range spatial features, and is computationally efficient. A formula for
estimates of long-range order spatial artifacts has also been proposed and
verified for the studied material system.

\paragraph{Acknowledgement.}
The authors acknowledge the Czech Science Foundation endowment under grants
Nos.~105/12/0331 (JN) and 105/11/0411 (AK and JZ). The work by JZ was also
partially supported by the European Regional Development Fund in the
IT4Innovations Centre of Excellence project (CZ.1.05/1.1.00/02.0070). In
addition, we would like to thank Ji\v{r}\'{i} \v{S}ejnoha and \mbox{Milan} Jir\'{a}sek
of CTU in Prague and anonymous referees for helpful comments on the manuscript.

\end{document}